\documentclass[prl,twocolumn,showpacs,superscriptaddress,amsmath,amssymb]{revtex4}

\usepackage{graphicx}%
\usepackage{dcolumn}%

\begin{document}

\title{"Kinks", Nodal Bilayer Splitting and Interband Scattering in YBa$_2$Cu$_3$O$_{6+x}$.}

\author{S. V. Borisenko}
\affiliation{Institute for Solid State Research, IFW-Dresden, P.O.Box 270116, D-01171 Dresden, Germany}

\author{A. A. Kordyuk}
\affiliation{Institute for Solid State Research, IFW-Dresden, P.O.Box 270116, D-01171 Dresden, Germany} \affiliation{Institute of Metal Physics of National
Academy of Sciences of Ukraine, 03142 Kyiv, Ukraine}

\author{V. Zabolotnyy}
\author{J. Geck}
\author{D. Inosov}
\author{A. Koitzsch}
\author{J. Fink}
\author{M. Knupfer}
\author{B. B\"uchner}
\affiliation{Institute for Solid State Research, IFW-Dresden, P.O.Box 270116, D-01171 Dresden, Germany}

\author{V. Hinkov}
\author{C. T. Lin}
\author{B. Keimer}
\address{Max-Planck Institut f\"ur Festk\"orperforschung, D-70569 Stuttgart, Germany}

\author{T. Wolf}
\address{Forschungszentrum Karlsruhe GmbH, Institut f\"ur Festk\"orperphysik, D-76344 Karlsruhe, Germany}

\author{S.~G.~Chiuzb{\u{a}}ian}
\author{L. Patthey}
\affiliation{Paul Scherrer Institut, CH-5232 Villigen PSI, Switzerland}

\author{R. Follath}
\affiliation{BESSY GmbH, Albert-Einstein-Strasse 15, 12489 Berlin, Germany}

\date{5 September 2005}
\begin{abstract}
We apply the new-generation ARPES methodology to the most widely studied cuprate superconductor YBa$_2$Cu$_3$O$_{6+x}$. Considering the nodal direction, we found noticeable
renormalization effects known as "kinks" both in the quasiparticle dispersion and scattering rate, the bilayer splitting and evidence for strong interband
scattering --- all the characteristic features of the nodal quasiparticles detected earlier in Bi$_2$Sr$_2$CaCu$_2$O$_{8+\delta}$. The typical energy scale and the doping dependence of the
"kinks" clearly point to their intimate relation with the spin-1 resonance seen in the neutron scattering experiments. Our findings strongly suggest a
universality of the electron dynamics in the bilayer superconducting cuprates and a dominating role of the spin-fluctuations in the formation of the
quasiparticles along the nodal direction.

\end{abstract}

 \maketitle

The presence of similar CuO planes in all high-temperature superconducting (HTSC) cuprates provides a key to understand their puzzling generic phase diagram
\cite{PhaseDiagram}. On the other hand, the sensitivity of the maximal transition temperatures ($T_c$) to the rest of the lattice implies that one should study
the cuprates as a class of materials aiming to gain the knowledge of how to drive $T_c$. Due to different purity of the samples, their mechanical properties,
surface quality, size of the available single crystals etc., a single experimental technique often provides information only about a few members of the large
cuprate family. A known example in HTSC research is the trio of the spectroscopic techniques which can be considered as complementary to each other: scanning
tunneling spectroscopy \cite{Davis}, inelastic neutron scattering (INS) \cite{VH1} and angle-resolved photoemission (ARPES) \cite{Damascelli}. The first one
prefers to deal with Bi$_2$Sr$_2$CaCu$_2$O$_{8+\delta}$ (BSCCO) and fails to study La$_{2-x}$Sr$_x$CuO$_4$ (LSCO) whereas the second one is dominantly applied to YBa$_2$Cu$_3$O$_{6+x}$ (YBCO) and LSCO. ARPES has been successfully applied to many hole-doped
cuprates. Though a hallmark of the modern ARPES is the sensitivity to the many-body effects and up to now only BSCCO and to a lesser degree LSCO turned out to
be suitable targets for such studies. Investigations of YBCO by means of photoemission, despite a rich history \cite{History}, have been virtually stopped on
the verge of the new millennium. In the last (to the best of our knowledge) report \cite{Lu} the authors do find a single many-body related feature, renown
"peak-dip-hump" structure, but leave open some fundamental questions, like the origin of the surface state and the presence of the bilayer splitting (BS).
Thus, even the underlying band structure of YBCO, a basic prerequisite to study many-body interactions by ARPES, remains unclear \cite{Damascelli}.

%%The reason for such an unusual situation in HTSC field  - surface quality, resolution and twinning at an earlier stage.
%%providing a natural bridge to the inelastic neutron scattering experiments.

In this Letter we extend the power of the new-generation ARPES to YBCO cuprates. Our first objective is the nodal direction which is known to be free of the
influence of the surface state \cite{History, Lu}. We find not only a clear indication for the existence of the bilayer splitting but also other typical
characteristics of the nodal quasiparticles found earlier in BSCCO and LSCO, such as dispersion anomalies dubbed "kinks", their doping dependence, and a scattering rate at low energies
consistent with the quadratic behaviour. Remarkably, we also found evidence for the interband scattering
which we have recently observed in a quite complicated experiment using circularly polarized light in BSCCO~\cite{InterbandBoris}.

ARPES measurements were made on two different beamlines, the U125/1-PGM beamline at BESSY and the SIS-beamline at the Swiss Light Source, using the same
portable end-station. Both beamlines deliver photons with comparable characteristics in terms of flux, resolution and polarization in the energy range from 50
to 55 eV. The overall energy resolution was set to 12 meV, and the angular resolution to 0.2$^\circ$. To ensure the generality of the observations, we have
measured several samples for each doping level on different synchrotrons.
%%In spite of the different beam-spot sizes the data are reproducible which implies a homogeneous
%%surface.
Data were taken on detwinned high-quality single crystals. These were synthesized by the solution-growth technique and after cutting, they were annealed to the
desired oxygen doping. Subsequently, they were detwinned by subjecting them to a uniaxial mechanical stress at elevated T \cite{Hinkov}. The samples thus
obtained exhibit a  $T_c$ = 90 K,  $\Delta T_c < $ 1~K (OD90K), $T_c$ = 61 K, $\Delta T_c < $ 2 K (UD61K) and $T_c$ = 35 K, $\Delta T_c < $ 3 K (UD35K). All
samples have exposed mirror-like surfaces after cleavage.

A typical and essential component of the ARPES experiment is the possibility to select a suitable excitation photon energy. This advantage is illustrated in
Fig.\,1 where we show raw intensity maps taken along the nodal direction (right panel inset) in the UD35K sample at 30 K as a function of photon energy. In
agreement with LDA calculations \cite{Andersen} one observes two features corresponding to the bonding (B) and antibonding (A) components of the electronic
structure of YBCO. Depending on the photon energy used, their intensity ratio changes leading to the dominance of the bonding band at h$\nu$~=~55~eV. This
behavior is indeed expected as the similar change of the photon energy ($\sim$ 5 eV) resulted in significant variation of the bonding/antibonding intensity
ratio in the case of BSCCO, where it was attributed to matrix element effects \cite{Kord_PRL02, Boris_PRL03}. This similarity brings an additional confidence
that the two features shown in Fig.\,1 are indeed bonding and antibonding states. Such a suppression of the antibonding band allows us to focus on the details
of the bonding spectral distribution, knowing that we deal with a single feature.

\begin{figure}[t!]
\includegraphics[width=8.47cm]{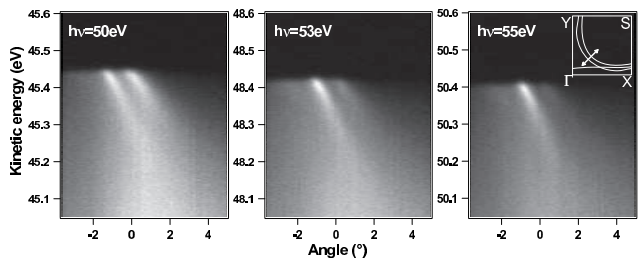}%
\caption{\label{First} Raw ARPES photoemission intensity measured using three different photon energies as a function of angle and kinetic energy of the
photoelectrons. The inset schematically shows the LDA-predicted Fermi surface and a cut in the k-space along which the data have been taken.}
\end{figure}

In Fig.\,2a we plot positions of the bonding MDC's maxima as a function of energy - a standard procedure to obtain the so called "experimental dispersion"
which offers a direct access to the real part of the self energy provided the bare band dispersion is known. The deviation of this dispersion from the straight
line passing through the Fermi momentum (dashed line in Fig.\,2a), seen here to be significant already at $\sim$ 50 meV binding energy, is the famous "kink"
detected before and well characterized in BSCCO and LSCO \cite{Damascelli, Kord_new}. There, the origin of this feature is still controversial and its
observation in a different cuprate can shed some light on this important issue. Upon closer consideration of the dispersion curve we have noticed the presence
of another, slightly less defined, "kink" at approximately 150 meV (grey arrow). This feature is less defined, because the dispersion above and below this
energy is not linear and the kink is seen only as a change of the curvature. The data shown in Fig.\,2b confirm the presence of both features. Here the width
of the MDC's, which is directly proportional to the imaginary part of the self energy (scattering rate), is plotted as a function of energy. As imaginary and
real parts of the self energy must be Kramers-Kronig related, one does expect that the characteristic features of one curve should reappear in the other. We
can tentatively assign 70 and 180 meV binding energies to the two scattering rate kinks. Not only the presence of the kinks is reproduced in the scattering
rate, but also the quadratic low energy behavior (dashed line in the right panel) agrees remarkably well with the linear dispersion near the Fermi level
clearly indicating the presence of quasiparticles \cite{Kord_PRB05}. In the insets we show a typical $k_F$-EDC and E$_F$-MDC in order to illustrate our high
resolution measurements.

\begin{figure}[t!]
\includegraphics[width=8.47cm]{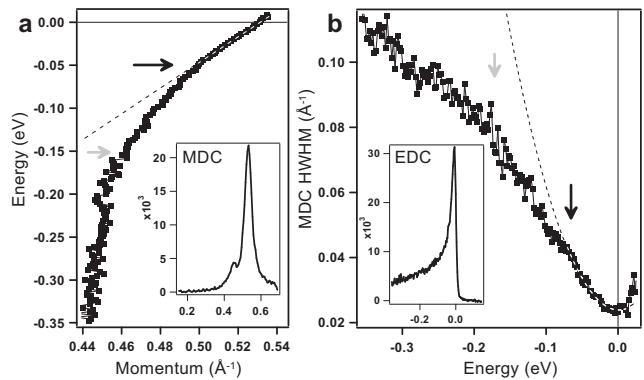}%
\caption{\label{Second} Positions of the MDC's maxima (a) and their widths (b) as a function of binding energy for the UD35K sample. Dashed curves are fits to
the low energy data. Black and grey arrows mark the energy positions of the low and high energy "kinks" respectively. Insets show exemplary MDC and EDC.}
\end{figure}

Fig.\,3 represents a collection of data taken in the series of differently doped YBCO single crystals. Examination of the mass enhancement as a function of
charge carrier concentration reveals a pronounced dependence of both low and high energy parts of the dispersion. Absolute values of the kink energy are
extracted in an approximate but very simple way. The low energy part of the data between E$_F$ and some gradually increasing energy is fitted to a line passing
through $k_F$. The high energy limit is called then a "kink energy" when the residuals start to sharply increase. As follows from the values given in Fig.\,3
there is a clear energy shift of the kink to higher binding energies upon doping. This is in agreement with the previous studies on BSCCO \cite{Johnson,
Gromko, Kord_new} and LSCO \cite{Kord_new}, though the effect is more pronounced, most likely because of the clear separation between bilayer split components.
We note that there exists also a systematic change of the Fermi velocity: 1.47 eV{\AA} (UD35K), 1.63 eV{\AA} (UD61K) and 1.64 eV{\AA} (OD90K).

Another familiar trend seen before in BSCCO \cite{Johnson} and LSCO \cite{Zhou_nature} is the change of the high energy slope of the dispersion. Upon doping it
becomes less steep. 
%%This is illustrated by the rightmost panel of Fig.\,3.
%%This behavior is possibly related to the existence of the second kink, which we discuss below.
The second kink, introduced in Fig.\,2 for the UD35K sample, is visible also in the UD61K sample, but shifted to lower energy ($\sim$ 120 meV). If the energy
of the second kink follows this trend further with doping, it is not surprising not to observe it in the slightly overdoped sample: most likely it merges with
the upper one.

%%At the same time there are peculiarities not clearly seen in earlier studies, but noticed only recently (gromko,kord, johnson?). Most likely this is because of
%%the clear separation between the bonding and antibonding bands.

\begin{figure}[t!]
\includegraphics[width=8.47cm]{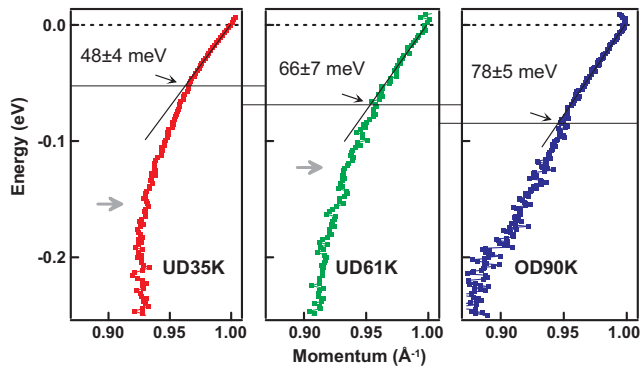}%
\caption{\label{Third} Dispersion as a function of doping. Horizontal lines mark the energies of the kink, where dispersion starts to deviate from the straight
line. Grey arrows show the position of the second, high energy kink. Momentum scales are matched by putting $k_F$=1 $\AA^{-1}$.}
\end{figure}

Tuning the excitation energy back to see both features, as in Fig.\,1a, we show in Fig.\,4a and 4b dispersions corresponding to the A and B bands. The given
values of the nodal BS in momentum units are nearly five times larger than those in BSCCO \cite{KordNBS}, though comparable in differently doped samples. We
take the opportunity to clearly distinguish between A and B bands to check another characteristic property of the near-nodal excitations in BSCCO --- the
recently observed strong interband scattering \cite{InterbandBoris}. To recreate similar experimental conditions we select a cut in the Brillouin zone which
makes 10$^\circ$ with the nodal direction. The result is shown in Fig.\,4c. This time there was no need to use circularly polarized light to separate bonding
from antibonding band. Already the raw data imply that the intensity distributions of the A and B features are different. MDC's taken at the Fermi level and at
80 meV binding energy (solid yellow curves in Fig.\,4c) both have a double-peak structure, but the relative intensity of the two components is significantly
changing. This is a typical signature of the effect, since the intensity is inversely proportional to the imaginary part of the self-energy and thus to the
scattering rate \cite{InterbandBoris}. Simply speaking, the more intense the MDC, the longer is the life of the corresponding quasiparticle. A more direct and
important observation of the different scattering rates is presented in Fig.\,4d. We have fitted MDC's like those shown in Fig.\,4c with two Lorentzians and
plotted their widths as a function of binding energy. There is an unambiguous crossing of the two curves at $\sim$ 50 meV making it qualitatively similar to
the BSCCO case and thus indicating that the interband scattering suggested earlier to explain the effect \cite{InterbandBoris} is plausible also here. The
interband scattering means that the hole made in the antibonding band will most probably be filled by the "bonding" electron and vice versa. Within this
scenario the difference in the scattering rates becomes natural: the densities of states (DOS) which correspond to the bonding and antibonding bands are
different --- both are peaked at the energies of the corresponding van Hove singularities. In the case of BSCCO a simple estimate has shown that they cross at
$\sim$ 100 meV, exactly where the scattering rates are equal. The data from Fig.\,4d are taken from the slightly overdoped sample where the antibonding band's
van Hove singularity lies closer to the Fermi level. In addition, the bilayer splitting in YBCO is larger than in BSCCO. Therefore the crossover energy of
$\sim$ 50 meV perfectly fits the mentioned scenario.

\begin{figure}[t!]
\includegraphics[width=8.47cm]{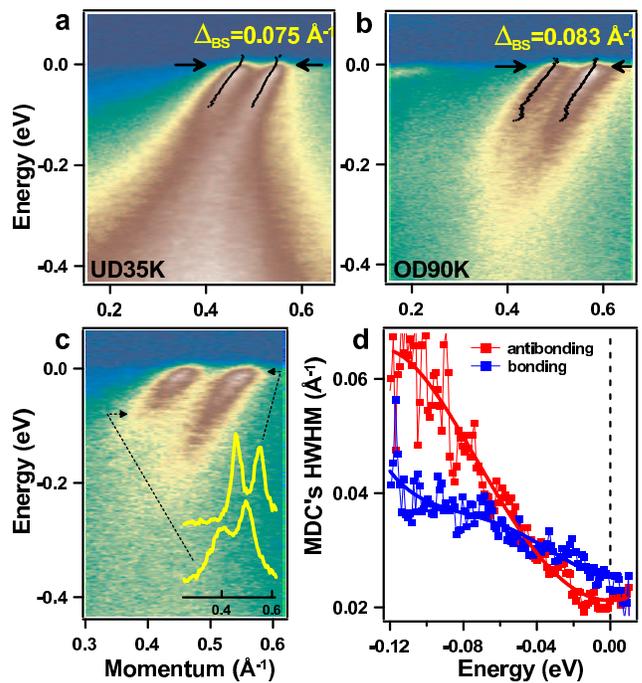}%
\caption{\label{Fourth} (a) and (b) Constancy of the nodal bilayer splitting. Solid lines are experimental dispersions. (c) Intensity map taken 10$^\circ$ away
from the nodal direction in the OD90K sample. Yellow curves are the MDC's corresponding to the binding energies marked by black arrows. (d) Widths of the MDC's
from the panel (c).}
\end{figure}

Now we briefly overview the obtained results: (i)the nodal bilayer splitting is present in a series of differently doped YB$_2$C$_3$O$_{6+x}$ single crystals
and is virtually independent on $x$; (ii) the experimental dispersion of the bonding band exhibits a kink, which systematically shifts to higher binding
energies upon increasing the hole concentration; (iii) there is another, less defined change of the curvature residing at higher energies which shifts towards
the upper kink and probably merges with it already in the slightly overdoped sample; (iv) scattering rates which correspond to the bonding and antibonding
bands are different being inversely proportional to the antibonding and bonding DOS respectively. While the result (i) is just a consequence of the underlying
band structure predicted by LDA calculations \cite{Andersen}, the other three are typical many body effects detected for the first time in YBCO. We have shown
earlier \cite{KordPRL04} that one can capture the main essentials of the electron dynamics along the nodal direction considering fermionic and bosonic channels
of scattering. We implement this approach also here.

We believe that it is the fermionic channel that is responsible for the second elusive kink. The real part of the self-energy goes to zero at the energy
related to the occupied bandwidth thus having a maximum at some lower energy which results in the kink \cite{Kord_new}. As the occupied bandwidth decreases
with doping, the second kink naturally follows this trend. Another possible consequence of this behavior is that the high energy velocity then decreases since
the bottom of the band slowly approaches the Fermi level. This effect is indeed observed here (Fig.\,3) and earlier in BSCCO \cite{Johnson} and LSCO
\cite{Zhou_nature}.

Results (ii) and (iv) can be understood in terms of the bosonic scattering channel. We do not repeat the arguments here as these features appear to be
universal and have been discussed before in relation to BSCCO and LSCO \cite{Damascelli,InterbandBoris}. What is controversial is the origin of the bosonic
mode itself. The specificity of the present result is that the spectra are taken on exactly the same samples which have been studied by inelastic neutron
scattering \cite{Hinkov}. Thus, there is a unique opportunity to compare directly the typical energy scales of the charge and spin dynamics in the identical
samples. As seen from Fig.\,3, the energy of the kink monotonically increases with doping. This is exactly the property of the resonance peak \cite{VH1}
measured on the same single crystals \cite{Hinkov, Hinkov_un}. The absolute energy of the kink does not match the energy of the resonance, and it actually should not as
the resonance couples effectively only those regions on the Fermi surface which are separated by the momentum vector ($\pi$, $\pi$) or close to them. Nodal
regions are coupled by the vector (0.8$\pi$, 0.8$\pi$). According to recent experiments \cite{Pailhes} and theoretical calculations \cite{Eremin} there is
another resonance in the magnetic spectrum near the vector (0.8$\pi$, 0.8$\pi$) at higher energies than the conventional ($\pi$, $\pi$)-resonance. This new
collective mode, being not fundamentally different from the resonant mode at ($\pi$, $\pi$), seems to track its doping dependence thus perfectly supporting our
observations. We do not think however that the nodal spectra can be fully explained in terms of a coupling to a single mode. The mentioned magnetic modes are
rather sharp in momentum space. In reality the experimental dispersion always smoothly evolves in {\bf k}-space. It is also known that not only resonant parts
of the magnetic excitation spectrum can be involved in the formation of the kinks, but also a spin-fluctuation continuum \cite{Chubukov}. It seems that more
rigorous calculations which take into account full integration over the Fermi surface are needed to reproduce quantitatively the experiment. We believe that a
phonon scenario is unlikely, as we are not aware of the existence of any phonon mode which scales with doping (and thus with $T_c$) in such a way.

The same is true for interband scattering. If our interpretation is correct, then the mediator of such scattering should have a corresponding symmetry. It is
known that the magnetic excitations are dominant in the odd with respect to the layers exchange within a bilayer channel \cite{VH4}. We are not aware of a
similar property of any phonon mode which can couple nodal electrons. On the other hand, the emergence of interband scattering as a universal property of the
cuprates can play a decisive role in the search for the mediator of such scattering and thus, perhaps, of the pairing. An interesting experiment would be to
demonstrate that the interband scattering is peculiar to a system where the coupling between electrons and phonons/spin fluctuations is {\it known} to be
dominant. At present, a multitude of experimental facts lead us to conclude that the spin fluctuations are the most likely candidates to mediate the pairing in
cuprates.

In conclusion, our data provide a first observation of the "kinks" in both dispersion and scattering rate, as well as interband scattering, in a bilayer
cuprate other than BSCCO. This implies a universality of these features which appear to be crucial for a complete understanding of HTSC.

The project is part of the Forschergruppe FOR538. This research project has been partially performed at the Swiss Light Source, Paul Scherrer Institut, Villigen, Switzerland and
has been  supported by the European Commission under the 6th Framework Programme through the Key Action: Strengthening the European Research Area, Research Infrastructures. Contract n$^\circ$: RII3-CT-2004-506008. We also thank R. H\"ubel for technical support.

\end{document}